\documentclass[runningheads]{llncs}
\usepackage[T1]{fontenc}
\usepackage{graphicx}
\usepackage{fancyvrb}
\usepackage{xcolor}
\usepackage{multicol}
\usepackage[inline]{enumitem}

\DefineVerbatimEnvironment{pseudocode}{Verbatim}{
    fontsize=\small,
    fontfamily=tt,
    xleftmargin=0cm,
    baselinestretch=1.1
}
\begin{document}
%
\title{Data Structures for Private Token Transfers in TEE-Based Networks}
%
\titlerunning{Data Structures for Private Token Transfers in TEE-Based Networks}
%
\author{Blake Regalia\inst{1} \and
Benjamin Adams\inst{2}\orcidID{0000-0002-1657-9809}}
\authorrunning{B. Regalia and B. Adams}
%
\institute{Solar Republic LLC, USA, 
\email{blake.regalia@gmail.com}\\
\and
Department of Computer Science and Software Engineering, University of Canterbury, New Zealand,
\email{benjamin.adams@canterbury.ac.nz}}
\maketitle              
\begin{abstract}
Trusted execution environment (TEE) based confidential smart contract networks promise privacy but remain vulnerable to storage access pattern attacks that can link senders and recipients in token transfers. When contracts update recipient balances during transfers, the unique storage keys accessed reveal transaction relationships even when data is encrypted. This paper introduces two novel data structures to address this vulnerability: the Delayed Write Buffer (DWB) and the Bitwise-Trie of Bucketed Entries (BTBE). The DWB delays recipient balance updates by buffering pending transfers and randomly settling entries, breaking the direct correlation between transfer execution and recipient storage access. The BTBE further enhances privacy by grouping addresses into constant-sized buckets, preventing flooding attacks and creating anonymity sets for balance queries. Additionally, we present a private notification system enabling real-time, privacy-preserving push notifications for confidential contracts. Our domain-specific approach leverages the unique characteristics of token transfers---asymmetric balance updates and tolerance for delayed settlement---to achieve practical performance with probabilistic anonymity guarantees. 

\keywords{Private transfer \and Smart contract \and Trusted execution environment.}
\end{abstract}
\section{Introduction}

Trusted execution environment (TEE) based confidential smart contract networks have emerged in recent years in recognition of the limitations posed by public blockchain databases \cite{li2022sok,smart2023computing}. These networks leverage hardware-based isolation to execute smart contract code securely, with the intention that sensitive data remains confidential from external observers and even from the network node operators. Programmable privacy provides application-level privacy, which provides a flexible set of privacy characteristics for users beyond simple transactional privacy \cite{benarroch2024sok}. Examples of blockchains that now implement programmable privacy with TEEs include Secret Network, Oasis, and Phala \cite{cheng2018ekiden,woetzel2016secret,yin2019phala}. Beyond these, TEE-based confidential features are increasingly being adopted by other natively public chains. TEE network confidentiality is touted as enabling a number of new types of privacy-preserving financial applications, given that TEEs are far more efficient than competing privacy technologies such as zero-knowledge proof systems (for two-party interaction) or fully homomorphic encryption.

For Secret Network and Oasis, data confidentiality is implemented in the database record through encrypted keys and values. This approach leaves applications such as token contracts susceptible to access-pattern attacks when they are implemented naively, e.g. via a simple translation of the ERC20 standard \cite{desai2021secauctee,jean2024sgxonerated}. When performing token transfers, secret contracts must store balances and transfer histories in their internal (encrypted) key-value databases. However, even though the database key is encrypted, it is unique to the recipient. This allows an attacker to monitor which database keys are read from and written to each execution to deduce storage access patterns and thus correlate transfers to a particular recipient the next time that key is read from or written to. Since the contract must access a sender's balance when executing an outgoing transfer (to make sure they have enough to cover the spend), this leads to a conundrum: how can we update a recipient's balance during a transfer without accessing a storage area associated with their account?

Jean et al. \cite{jean2024sgxonerated} demonstrated a catastrophic privacy failure that links senders and recipients for token transfers using Secret Network's SNIP-20 token standard, and they suggested two ways to mitigate the issue. Their first proposed solution was to use \textit{decoys}, essentially a set of user-supplied mock recipients for a given transaction. This solution was implemented for Secret Network tokens soon after the exploit became known. However, because it relies on user intervention and is applied sparingly and inconsistently by dApps and users, it has largely failed to achieve the desired result. The second proposed solution was to implement an oblivious RAM (ORAM) implementation, either at the network-level, overriding get/set operations for the encrypted database, or within the contract \cite{goldreich1987towards}. However, issues around the efficiency and best way to develop that remain to be explored. While the best-known ORAM schemes achieve $O(\log N)$ complexity per access, much of ORAM’s cost comes from bandwidth blowup (reading and writing more data than necessary) potentially gas-prohibitive in a blockchain context \cite{stefanov2013path}. Other solutions that might include off-chain computation require multiple-steps and would not allow transactions to occur per block. 

In this paper, we present a third, domain-specific option using two new data structure constructions: the Delayed Write Buffer (DWB) and the Bitwise-Trie of Bucketed Entries (BTBE), which provide unlinkability between senders and recipients for token transfers with a low-complexity design that can be implemented within the token contract. In addition, we complement this functionality with a specification for private notifications, enabling immediate, private feedback to recipients of token transfers. While this solution does not have the full generality of ORAM, when tailored to the narrow case of token transfer storage, it has minimal and consistent gas/computation overhead and provides probabilistic anonymity.

\subsection{Related work}

\subsubsection{Oblivious RAM.}

The concept of Oblivious Random Access Memory (ORAM) was first introduced by Goldreich and Ostrovsky almost 30 years ago as a cryptographic primitive designed to hide memory access patterns from adversaries who can observe which memory locations are accessed during program execution \cite{goldreich1996software}. The fundamental security guarantee of ORAM is that the sequence of memory accesses produced by any two programs with the same running time are computationally indistinguishable to an observer monitoring the physical memory interface. Formally, an ORAM scheme provides oblivious simulation of a program's memory access pattern. Given a sequence of memory operations (read/write requests to logical addresses), an ORAM construction transforms these into a sequence of physical memory accesses such that the physical access pattern reveals no information about the logical access pattern, beyond the total number of operations performed. This property is crucial in scenarios where an adversary can monitor memory access patterns but cannot observe the actual data being accessed, such as in cloud computing environments or when using untrusted storage systems.

Significant advances in ORAM efficiency emerged with tree-based constructions, most notably Path ORAM introduced by Stefanov et al. \cite{stefanov2013path}. Path ORAM organizes data in a binary tree structure where each data block is assigned to a random leaf, and accessing a block requires reading and writing an entire root-to-leaf path. This approach achieves $O(\log N)$ overhead per access while maintaining strong security guarantees through periodic reshuffling of data along accessed paths. Circuit ORAM further optimized tree-based constructions to achieve asymptotically optimal $O(\log N)$ overhead with smaller constants \cite{wang2015circuit}. 

Traditional ORAM constructions assume that both read and write operations must be performed obliviously. However, many practical applications require only write privacy. Roche et al. introduced deterministic, stash-free write-only ORAM (WO-ORAM), demonstrating that significant performance improvements are possible when read operations do not require obfuscation \cite{roche2017deterministic}. WO-ORAM constructions typically employ a delayed write strategy where write operations are buffered and processed in batches, rather than immediately updating the target memory location. This approach allows multiple write operations to be combined and randomized before being committed to storage, breaking the correlation between the timing of write requests and physical storage locations.

Recognizing that real-world applications often exhibit repeated access patterns to the same memory locations, Dautrich et al. introduced Burst ORAM to minimize response times for such access patterns \cite{dautrich2014burst}.  Burst ORAM's batching strategy demonstrates how application-specific optimizations can be incorporated into ORAM designs while maintaining security guarantees. The scheme ensures that an adversary observing the physical access pattern cannot determine whether a given operation resulted from a single logical access or from multiple accumulated accesses to the same location.

\subsubsection{Access-pattern and side-channel vulnerabilities.}

The importance of hiding storage access patterns extends beyond traditional memory systems to encrypted databases and distributed storage systems. Islam et al. demonstrated that even when database contents are encrypted, access pattern leakage can compromise user privacy by revealing which records are being queried \cite{islam2012access}. Their work showed that statistical analysis of access patterns can lead to significant information leakage, even in the absence of plaintext data. This analysis is particularly relevant to programmable privacy blockchain systems, where transaction data may be encrypted but storage access patterns during smart contract execution can reveal information about user interactions. The challenge is compounded in blockchain systems because the execution environment may be partially trusted or controlled by potentially malicious node operators.


The effectiveness of privacy-preserving storage systems depends on their resistance to side-channel attacks, particularly timing-based attacks that can reveal information about internal system state. Kocher et al. demonstrated that variations in execution time can leak cryptographic secrets, establishing the importance of constant-time algorithm design \cite{kocher1999differential}. Similarly, Bernstein showed how cache timing variations can compromise the security of cryptographic implementations \cite{bernstein2005cache}.



\subsection{Contribution}

While existing ORAM constructions provide strong theoretical foundations, their direct application to blockchain token transfers presents significant efficiency and practicality challenges. Traditional ORAM schemes are designed for general-purpose memory access patterns and do not exploit the specific characteristics of token transfer operations, such as the asymmetric nature of balance updates (one decrease, one increase per transfer) and the tolerance for delayed settlement of incoming transfers. Furthermore, existing approaches do not address the unique attack vectors present in blockchain systems, such as flush attacks where adversaries can manipulate buffer contents through coordinated transactions, or the need to maintain identical gas consumption patterns across different execution paths to prevent side-channel leakage.

The domain-specific nature of token transfers enables several novel optimizations not possible in general ORAM constructions. First, the unidirectional flow of value allows for asymmetric treatment of sender and recipient operations. Senders must be processed immediately for balance verification, while recipients can tolerate delayed processing through buffering mechanisms. Second, the append-only nature of transaction histories permits the use of linked-list accumulation strategies that eliminate the need for complex data reshuffling while maintaining privacy guarantees. Third, the finite and predictable set of operations (balance queries, transfers, and history queries) allows for specialized constant-time algorithms that are more efficient than general-purpose oblivious data structures. These domain-specific insights enable a hybrid approach that combines write-only ORAM buffering with tree-based bucketing and novel anti-flush mechanisms, achieving practical performance.

\section{Requirements}

To address the challenges identified, we articulate a set of design goals and security properties. These capture the constraints under which the system must operate and the guarantees it must provide. This approach allows us to iteratively specify necessary components, avoid redundancies, and ensure that solutions do not conflict with one another. \\

\begin{enumerate}[label=, leftmargin=*, nosep]
\item \textbf{R1. Storage Access Patterns $\Rightarrow$ Transfers Table} \\
\textit{Property:} Transfers must not produce correlated storage access patterns linking sender and recipient. \\
\textit{Mechanism:} Instead of writing to the recipient’s storage directly, write pending transfers to a table that gets fully read and overwritten on each execution.
\\

\item \textbf{R2. Finite Table Capacity $\Rightarrow$ Delayed Write Buffer} \\
\textit{Property:} The transfer table has finite capacity; when full, it must be updated without leaking information about which entries are evicted.
\\
\textit{Mechanism:} Randomly select an entry from the table (that is, the buffer) to evict and update the stored balance of its recipient (i.e., ``settle'' the entry).
\\

\item \textbf{R3. Flush Attacks $\Rightarrow$ Accumulation} \\
\textit{Property:} Repeated transfers to the same recipient must not reduce the anonymity set or reveal prior recipients through forced settlement, as this would allow an attacker to ``flush'' the rest of the buffer and record which keys are accessed.
\\
\textit{Mechanism:} Transfers to recipients who have an existing entry accumulate in place rather than creating new ones or settling and replacing existing ones.
\\

\item \textbf{R4. Side Channel Leaks $\Rightarrow$ Constant Time Operations} \\
\textit{Property:} Execution time, storage access timing, and gas usage must remain consistent and independent of buffer state, preventing leakage to a malicious observer.
\\
\textit{Mechanism:} All operations before the final storage read are constant-time, independent of transfer parameters.
\\

\item \textbf{R5. Storage Write Leaks $\Rightarrow$ Phony Writes} \\

\item \textbf{R5a. Recipient Entry} \\
\textit{Property:} Due to the Accumulation mechanism in R3, a random entry does not need to be evicted if the transfer recipient already has an entry. However, this would lead to a branch in execution where sometimes an entry gets settled and sometimes not. More importantly, the absence of a write would leak that the recipient of the transfer was already in the buffer.
\\
\textit{Mechanism:} If the recipient is already in the buffer, rather than settling their existing entry, perform a phony write to an unrelated address's storage location.
\\

\item \textbf{R5b. Owner/Sender Balance} \\
\textit{Property:} Outgoing transfers may exceed the value of an owner's \textit{stored} balance, but not when combined with the cumulative sum of their incoming transfer entries in the buffer (that is, their \textit{total} balance). Checking and updating an owner's total balance must not reveal whether they had an entry in the buffer.
\\
\textit{Mechanism:} The contract always checks and settles the owner's buffer entry if it exists. If the owner does not have an entry in the buffer, in order to produce the same storage write pattern, it performs a phony write to the storage area of an unrelated address.
\\

\item \textbf{R6. Flooding Attacks $\Rightarrow$ Buckets} \\
\textit{Property:} A vigilant attacker with endless funds must not be able to deduce the entries in the buffer through a series of sandwiched flooding attacks. By monopolizing the buffer with transfers to known recipients, an attacker may establish a high degree of certainty over the buffer's contents. Albeit prohibitively expensive to perform at scale, a determined attacker may be able to correlate or de-anonymize an individual transfer event to a pre-selected victim recipient.
\\
\textit{Mechanism:} Instead of storing each owner's/sender's/recipient's balance and history under a single storage key, balances and histories are grouped into buckets of finite anonymity sets. Reads and writes occur at the bucket level, concealing individual access.
\\

\end{enumerate}

\section{Delayed Write Buffer (DWB)}

We introduce the Delayed Write Buffer (DWB), a domain-specific fixed-width data structure that gets fully read and overwritten on every transfer execution. A token contract requires only a single DWB instance to handle all token transfers. DWB contents are encrypted at rest in the contract's key-value store by virtue of the confidential contract platform, where it is stored under a constant storage key.
In naive implementations, the storage key associated with a recipient's balance is accessed during transfer executions to update their balance, revealing storage access patterns. With the DWB, instead of accessing any storage areas associated with the recipient, an entry containing the recipient's address, pending token amount, and a pointer to a linked list of transfer events is inserted into the DWB. A transfer event contains metadata such as the sender's address, date time, and an optional memo. Once the DWB has reached saturation, that is, every slot in its fixed-width capacity is occupied by an entry, the contract selects an entry from the buffer at random to ``settle'' in order to make room for the new entry. 
In order to prevent an attacker from deducing which entry was settled, the source of randomness must be private. In our implementation, randomness is provided by Secret VRF which uses the network's internal private key to derive a unique secure seed for RNG on every contract execution within the TEE. Settling an entry updates the associated recipient's stored balance and removes the entry from the buffer. This mechanism artificially delays the eventual write to a recipient's storage area by some random number of executions, avoiding direct storage access association between sender and recipient.

\subsection{Executing the transfer}

\begin{figure}
\centering
\includegraphics[width=\textwidth]{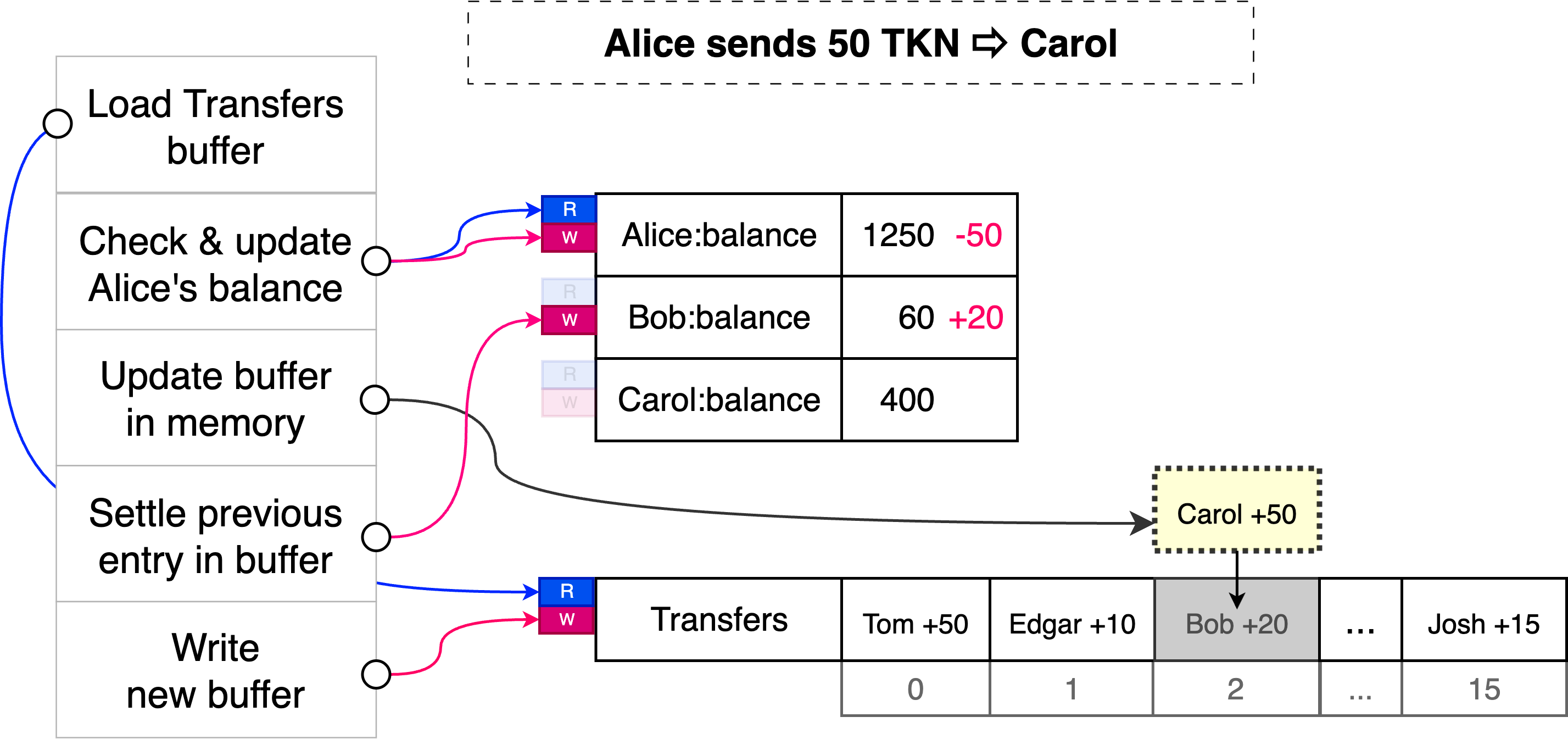}
\caption{An example transaction using a DWB, showing the order of read/write storage access operations. Notice how Bob's entry was randomly selected from the buffer to be settled in order to insert the new entry for Carol.} \label{fig:dwb}
\end{figure}

Here we describe an example (illustrated in Figure~\ref{fig:dwb}) of executing a transfer transaction using a DWB.

First, a new transfer event is saved to storage, keyed by a globally unique transfer event ID. Next, the contract loads the entire DWB from storage, selects an entry at random, and settles it.


At this point in the example process, nothing associated with the recipient's storage areas has been accessed. Instead, the sender of the transaction has incidentally updated the stored balance of a random account that was designated by an entry in the DWB, even though the transaction sender and the accessed account likely have no affiliation. This strategy greatly reduces the ability for an attacker to correlate sender and recipient through storage access patterns.

\subsection{Querying for balance and history}
\label{subsec:querying}

Entries stored in the DWB count towards a user's total balance. When Carol queries for her balance, the contract must search the DWB to find any entries where Carol is the recipient. It adds this value to her stored balance to arrive at her actual total balance. Queries for holistic transfer history must also access data from both storage areas (more on history below).

Our approach does not guarantee privacy during query operations. As opposed to transactions which are broadcasted to the entire network, queries are ostensibly private between the client and the node serving their RPC call. While the node cannot decrypt the query nor its outputs, and its execution is run within the TEE, a malicious node operator can monitor which storage keys are accessed from outside the enclave. For example, a victim repeatedly querying for their balance may reveal their balance storage key. If the victim also broadcasts transaction requests to the node, the attacker may be able to infer the client's account address and by extension their balance storage key. However, this information is less meaningful in practice when the DWB is used since each transfer execution does not access the recipient's balance storage key, leaving an attacker with a large anonymity set of potential senders to correlate with a given recipient. Additionally, as explained in Section \ref{sec:btbe}, a complementary data structure for stored balances further mitigates against these types of attacks by creating anonymity sets for stored balances.

\subsection{Owner's balance}

Executing a transfer requires a balance check of the owners's\footnote{The term \textit{owner} is used to distinguish the holder of funds from the message \textit{sender}, who may be executing on the owner's behalf via permissioned allowance.} account. Since a transfer must decrease the owner's total balance, the contract first settles any pending entries in the DWB designated for their account. In order to prevent leaking information about DWB contents, the owner's balance is overwritten regardless of whether or not such an entry is found. This action produces a definite storage access signal at the owner's stored balance key, establishing a likely association between the public message sender and the private account owner's storage area. However, as mentioned in Section \ref{subsec:querying}, this information is less meaningful to an attacker in practice, and is rendered even less so with the addition of the complementary data structure introduced in Section \ref{sec:btbe}.



\subsection{Distinct recipients}

A nontrivial privacy vulnerability arises when recipients are allowed to appear multiple times in the DWB. An attacker can \textit{flush} access to a victim's stored balance by executing many simultaneous transfers to them. Effectively, this would create a high probability that every entry in the DWB is designated to the victim, and would eventually reveal repeated access to the same storage key. To mitigate this, our implementation enforces that entries in the DWB are distinct by recipient address, and repeated transfers to the same recipient accumulate in the existing entry. In other words, if the recipient already has an entry designated to them in the DWB, the contract updates this entry with the cumulative transfer amount rather than inserting a new entry.



\subsection{Transaction history events}

\begin{figure}
\centering
\includegraphics[width=0.6\textwidth]{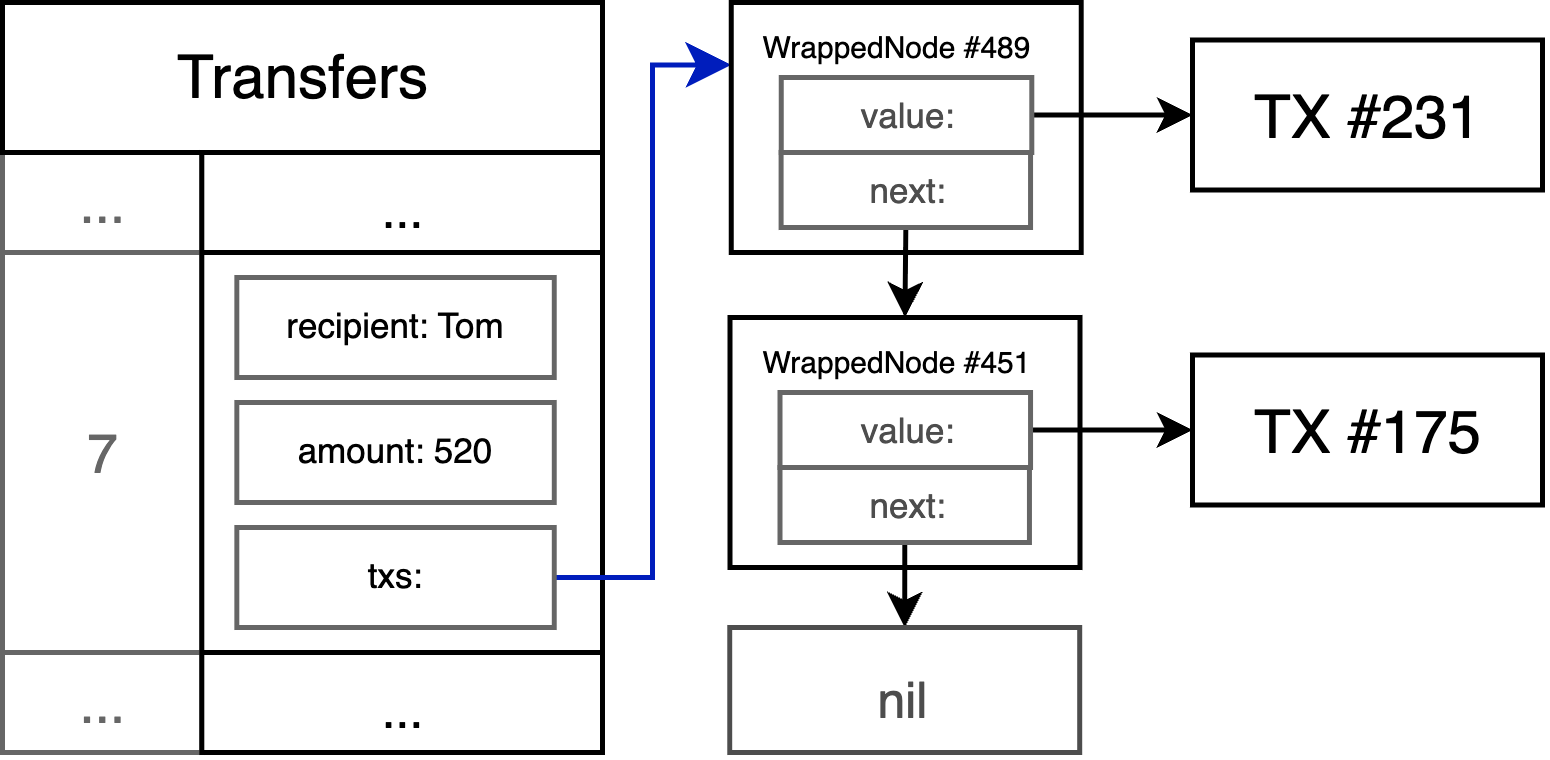}
\caption{Repeated transfers to the same recipient accumulate in a linked list of transaction events per DWB entry. Notice how insertions do not dereference previous items in the list. Thus, updating the list does not produce storage access patterns.} \label{fig:dwb-ll}
\end{figure}

Accumulating transfers in the DWB presents a challenge to storing historical records. In classic Secret tokens, users have the ability to query for their transaction history which includes all outgoing and incoming transfers. If the contract were to simply write each historical record to a storage area associated with the recipient, it would produce a storage access pattern and defeat the purpose of the DWB. Instead, each transfer event is appended to a global list and a reference to its ID is saved under a field in the DWB entry. Since repeated transfers to a given recipient accumulate in the DWB, the entry needs a way to store multiple events. Our solution is to use a linked list, where each new transfer inserts at the head of the list as shown in Figure~\ref{fig:dwb-ll}.

With this approach, an entry in the DWB can accumulate events ad infinitum. The storage areas associated with transaction history events are only ever accessed a single time across all executions (when they are written to). The only other time they are dereferenced is when they are read by the query node during private user queries for their transaction history.

\subsection{Saturation }

The DWB privacy mechanism relies on settling authentic entries to the storage area. During the first $k$ interactions, the DWB has empty slots that would not produce the desired storage access effects if settled, due to the lower anonymity set size. Therefore, DWB privacy is most effective once it is fully saturated. Our implementation tracks the saturation of the DWB during these initial executions and refrains from settling entries until it reaches full saturation.



\subsection{Data structure example for token transfers} 

The DWB is a simple byte sequence of $k$ entries, concatenated with a single \texttt{uint<$log_2(k)$>} which counts down the unused capacity while under-saturated. Each entry consists of the recipient, amount, a pointer to the head of the events linked list, and the current length of the list.

\begin{pseudocode}
20 bytes: recipient (canonical address)
+8 bytes: amount (uint64)
+5 bytes: events list ID (fits in uint64)
+2 bytes: list length
---------------------
= 35 bytes per entry
\end{pseudocode}

\subsection{Constant time search}

All operations prior to the final storage area write must run in constant time in order to prevent leaking information about branch conditions through side channels. This includes byte slice comparisons and the algorithm used to search for a matching owner address in the DWB during the transfer process. To give an example without constant time search, and assuming the entries were sorted by address, a hypothetical attack could brute force crafted addresses that probe insertion timing and use a binary search strategy to deduce the leading bytes of an address belonging to an entry in the DWB.





    

\subsection{Selecting buffer parameters}

Let $k$ be the capacity of the buffer. The probability that an entry for a given recipient has settled and thus accessed their stored balance after $n$ subsequent transfer executions is given by:

\[
P\left(\textrm{transfer settling}\right) = 1- \left( \frac{k - 1}{k} \right)^{n}
\]


\begin{figure}
\includegraphics[width=\textwidth]{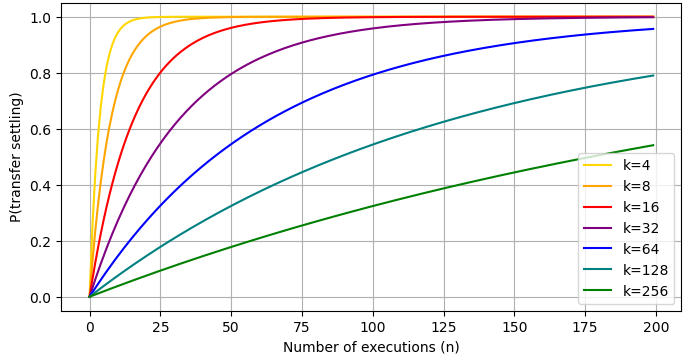}
\caption{Probability a transfer has settled, and thus the recipient's storage area has been accessed, after $n$ subsequent transfer executions.} \label{fig:dwb-access}
\end{figure}

We visualize this formula for various values of $k$ in Figure~\ref{fig:dwb-access}. Interpreting the diagram, for $k = 64$, we see an 80\% chance that an entry has settled after 100 executions. This uncertainty is what prevents an attacker from identifying which recent transaction actually transferred tokens to a given recipient.

Another way to interpret this formula is to answer at what point an attacker reaches high confidence that a victim's stored balance area was accessed within the last $n$ executions. In other words, for a buffer width of $k = 64$, reaching a 99.5\% confidence threshold is roughly equivalent to an anonymity set of size $336$.

\[
P(\textrm{transfer settling}) = 1 - \left(\frac{64-1}{64}\right)^n = 0.995
\]

\[
n = \frac{\ln(0.005)}{\ln(63/64)} = 336.4362...
\]

In physics, a $5\sigma$ threshold (5 standard deviations) is used as a statistical significance measure to 
indicate that an observed effect departs from the null hypothesis, which corresponds to a confidence level of $99.99994\%$. Reaching this confidence level for $k = 64$ would be equivalent to an anonymity set of size $909$.

\section{Bitwise-Trie of Bucketed Entries (BTBE)}
\label{sec:btbe}

We introduce a secondary data structure to improve the privacy of accessing users' stored balances during token transfers, complementing the benefits described above from the DWB. The Bitwise-Trie of Bucketed Entries (BTBE) manages a tree of constant-size tables which deterministically groups items by the leading bits of the cryptographic hash of their account address, better known as \textit{buckets} \cite{morrison1968patricia}. Grouping multiple records into a bucket creates a finite open anonymity set that limits the granularity of storage access patterns. To preserve privacy, the hash is derived using a secret key internal to the contract. The value of any bucket item stores the user's balance and transfer history.

Since the buckets have fixed capacity, new buckets must be created and existing ones must occasionally be rebalanced on insertion. The bitwise-trie component of the data structure enables efficient access to buckets, ensuring that the query and execution cost of locating and inserting entries grows logarithmically with new recipients. The height of the trie corresponds with the $i$th leading bit of the hash.

\begin{figure}
\includegraphics[width=\textwidth]{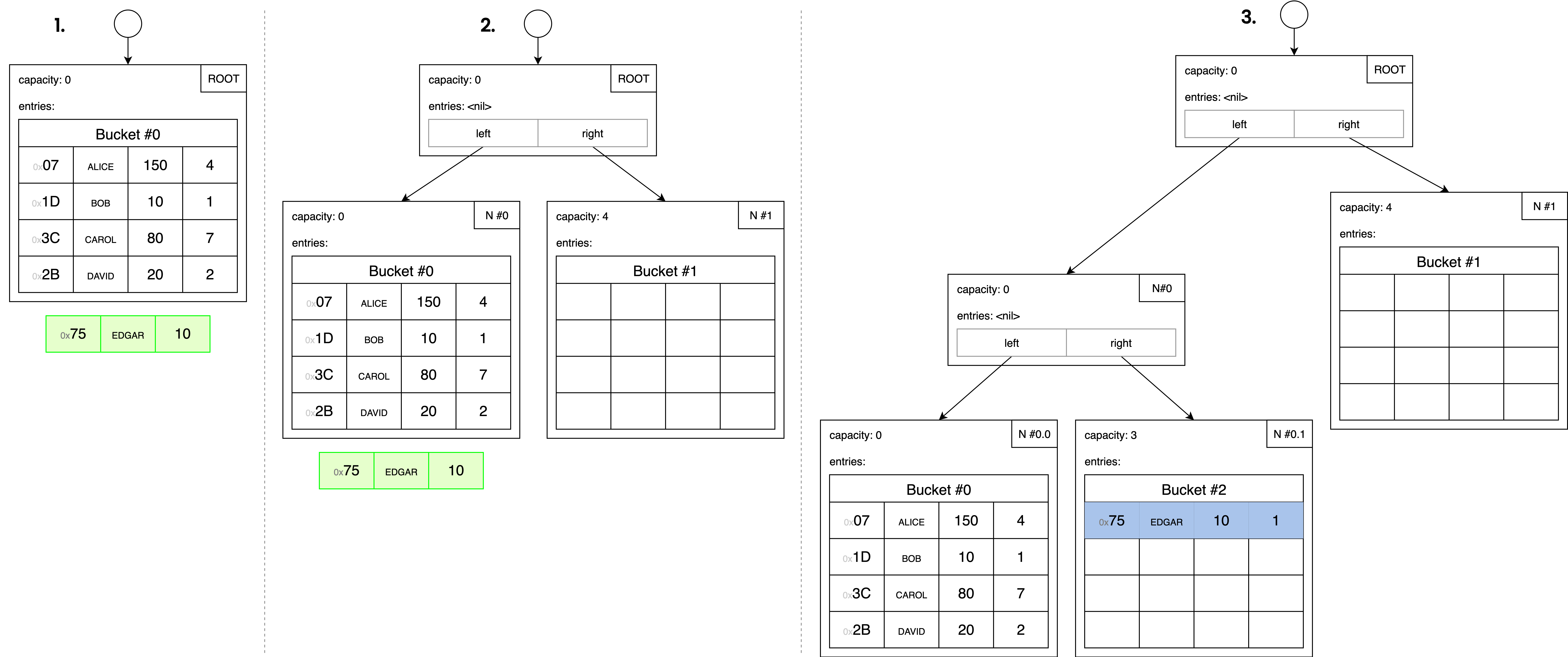}
\caption{Example of inserting a new entry in the BTBE. \textbf{1.} Bucket \#0 is full, resulting in a leaf node split. \textbf{2.} The inserted entry's hash routes it to Bucket \#0, which is again full, resulting in another leaf node split. \textbf{3.} The inserted entry's hash routes it to Bucket \#2, which has capacity and accepts the entry, terminating the insertion process.} 
\label{fig:btbe}
\end{figure}

The diagram shown in Figure~\ref{fig:btbe} illustrates the insertion of a new entry which requires rebalancing the trie. The example inserts an entry for Edgar's stored balance. His address hashes to the hexadecimal value \texttt{0x75 (0111 0101)}. Notice how the trie changes as new branches are created until reaching the 2nd leading bit of the hash, where Edgar's entry eventually finds capacity in Bucket \#2.

\section{Private notifications}\label{sec:notif}

Observing state changes to the encrypted databases of confidential smart contracts that affect an interested user is more difficult than the equivalent in a public blockchain. Wallets and dApps currently resort to a polling-based approach in order to notice changes to a user's private state within a contract. For example, a dApp might periodically query a set of token contracts to discover a new incoming transfer. However, this approach of querying contracts every so often is inefficient and can create unwanted load on query nodes. Additionally, there is no clear best practice for determining an optimal polling rate.

Here, we describe a specification that addresses these limitations by introducing a privacy-preserving push notification system for confidential smart contracts on Secret Network. The system enables clients to receive real-time notifications for specific events while maintaining complete privacy of recipient and contents. This approach leverages the existing Tendermint event infrastructure combined with cryptographic techniques to create globally unique notification identifiers only decipherable by the intended recipients.

Tendermint, the consensus engine for Secret Network, includes a publish-subscribe event stack that allows nodes to transmit network events directly to subscribed clients \cite{eugster2003many}. It does so using JSONRPC over WebSockets \cite{buchman2016tendermint}. Contracts are able to emit arbitrary plaintext data into an event log which gets broadcast by the aforementioned stack.

\subsection{Notification Framework}

The notification system operates on several key principles. Smart contracts generate globally unique, single-use notification identifiers using cryptographic hash functions. These identifiers serve as attribute keys in transaction logs, creating a discrete signaling mechanism visible only to intended recipients. The system distinguishes between different event types through channels, allowing contracts to separate notifications for transfers, messages, gaming events, and other application-specific activities.

The architecture relies on shared secrets between clients and contracts, termed notification seeds. These seeds serve as cryptographic key material to generate notification identifiers. Contracts derive default seeds using an internal secret combined with recipient addresses, enabling clients to obtain notification identifiers without executing transactions. For enhanced security, clients can establish custom seeds through cryptographic signatures.

\subsection{Channel Operating Modes}

The system supports three distinct operating modes, each optimized for different use cases and security requirements.

\begin{enumerate}[label=, leftmargin=*, itemsep=0.5em, topsep=2pt]
    \item \textbf{Counter Mode} provides the most convenient client experience by using sequential counters to generate unique notification identifiers. Clients need only recompute identifiers when receiving notifications and can search transaction history for missed events. 
    \item \textbf{TxHash Mode} eliminates side-channel vulnerabilities by incorporating transaction hashes into identifier generation. This approach provides stronger security guarantees but requires clients to recompute identifiers for every contract execution, increasing computational overhead.
    \item \textbf{Bloom Mode} enables efficient notification delivery to multiple recipients simultaneously. It uses Bloom filters, probabilistic data structures that encode set membership with controlled false positive rates to indicate which recipients should check for notifications \cite{bloom1970space}. The system employs constant-size filters with configurable parameters to balance efficiency and privacy requirements.
\end{enumerate}

\subsection{Information Hiding Techniques}

The system employs multiple layers of privacy protection to prevent information leakage. Notification data encryption uses the ChaCha20-Poly1305 authenticated encryption algorithm, chosen for its efficiency in constrained blockchain environments and widespread implementation support \cite{bernstein2005poly1305}.

To prevent traffic analysis attacks, contracts implement two critical privacy measures. First, all notification data is padded to constant lengths within each channel, preventing observers from inferring message content based on size variations. Second, contracts emit consistent numbers of log attributes regardless of actual notification activity, using decoy notifications to mask genuine events.

\subsection{Cryptographic Security Model}

The security model relies on established cryptographic primitives. HKDF (HMAC-based Key Derivation Function) generates deterministic yet unpredictable seeds from contract internal secrets and recipient addresses \cite{krawczyk2010cryptographic}. HMAC-SHA256 produces notification identifiers that appear random to external observers but remain deterministic for authorized parties \cite{bellare1996keying}. The secp256k1 elliptic curve signature scheme enables clients to establish custom shared secrets through cryptographic proof of key ownership \cite{brown2010sec2}. The security model relies on standard cryptographic assumptions including the discrete logarithm problem's hardness in secp256k1, HMAC-SHA256's pseudorandomness properties, and ChaCha20-Poly1305's semantic security \cite{bellare1996keying,brown2010sec2,bernstein2005poly1305}. These assumptions align with widely-used blockchain security models and benefit from extensive cryptographic analysis.




\subsection{Threat Model Considerations}

Counter Mode exposes a potential privacy vulnerability in which a malicious query node might associate a notification ID filter with a user's IP address. TxHash Mode eliminates this attack vector, though at the cost of increased client complexity. The system's privacy guarantees depend on proper implementation of constant-time operations and consistent event emission patterns. Contracts that fail to maintain consistent log sizes or data padding may leak information about notification patterns or recipient identities through traffic analysis.

For multi-recipient channels, proper Bloom filter parameter selection critically affects both efficiency and privacy. The filter size $(m)$, hash function count $(k)$, and underlying hash function $(h)$ must be chosen based on expected recipient group sizes and acceptable false positive rates. The specification recommends cryptographically secure hash functions to prevent preimage attacks while ensuring uniform distribution for filter effectiveness.


\section{Results and Discussion}

To evaluate our approach, we implemented a full reference implementation of a token contract on Secret Network incorporating the Delayed Write Buffer (DWB), the Bitwise-Trie of Bucketed Entries (BTBE), and the private notification system. The codebase is available for review here: \\ \url{https://anonymous.4open.science/r/snip20-reference-impl-F846/README.md}.

\subsection{Deployment and Adoption}

Our contract maintains the same Secret token interface for queries and executions, ensuring backwards compatibility with existing dApps. Since upgrading a live contract’s database schema is costly and complicated, we implemented a lazy migration mechanism for legacy balances. This enabled the community to upgrade major bridged assets (USDC, USDT, BTC, ETH, and sSCRT) to our privacy-preserving design. In total, 42 mainnet Secret tokens were upgraded, securing over USD~10~million.

\subsection{Gas and Execution Overheads}

We benchmarked the contract under a variety of workloads to measure execution overhead relative to the naive token contract baselines. The results show that:
\begin{enumerate*}[label=(\arabic*), itemjoin={{; }}, itemjoin*={{; and }}]
\item The DWB incurs a fixed gas overhead proportional to the buffer width $k$ during transfer executions
\item The BTBE incurs a variable gas overhead proportional to the height of the trie and size of the bucket
\item The notifications add a relative small fixed gas overhead to each execution thanks in part to the lightweight encryption algorithm ChaCha20-Poly1305.
\end{enumerate*}
In the worst case observed, gas usage increased by 26\% compared to the baseline SNIP-20 token. We consider this a modest tradeoff for the privacy protections gained.

\subsection{Privacy Guarantees in Practice}

Our experiments demonstrate that the DWB effectively decouples sender and recipient storage accesses. With a buffer capacity of $k = 64$, an attacker would require nearly 300 subsequent transactions to achieve even a 99\% confidence that a victim's balance has been accessed by one of those transfers. This simulates a large anonymity set for each recipient using only a fraction of the equivalent space, and substantially improves the privacy compared to naive contracts where association between sender and recipient are immediate due to direct storage access.
Additionally, the BTBE further obfuscates storage access patterns for user balances by grouping balances into buckets of constant size.

\subsection{Private Notifications}

The push-based notification system demonstrates significant improvements over polling. Clients receive real-time transfer alerts without exposing recipient identity or event timing to the network. In particular, the Bloom filter mode provides efficient multi-recipient signaling with false positive rates kept below 1\% under practical configurations. While TxHash mode increases client computation, it effectively mitigates side-channel risks, providing a tunable tradeoff between security and efficiency.

\subsection{Limitations and Future Work}

Our approach does not eliminate all forms of leakage. Query-based attacks remain a partial vector, since repeated balance queries by a malicious node operator can still reveal a user’s balance storage key. Similarly, DWB anonymity sets depend on transaction volume; during low-activity periods, settlement intervals shrink and anonymity guarantees weaken. Future work may explore adaptive buffer resizing, integration with network-level ORAM, and hybrid models that combine DWB/BTBE with zk-SNARK-based auditing for stronger guarantees.

\bibliographystyle{splncs04}
\bibliography{references}

\appendix

\section{Private notification algorithms}

\subsection{Contract internal secret derivation}

Contract initialization must establish an internal secret with high entropy, unknown to all parties including contract administrators.

\begin{pseudocode}
fun initializeContract(msg, env) {
  // gather entropy from sender
  let userEntropy := msg.entropy

  // extend entropy with environmental information
  let entropy := concat(
    env.blockHeight,
    env.blockTime,
    env.senderAddress,
    userEntropy
  )

  // the crux: obtain a unique, cryptographically-strong random value 
  // associated with this execution
  let seed := env.random()

  // very important: derive the contract's internal secret using HKDF
  let internalSecret := hkdf_sha256(
    ikm=seed, 
    salt=sha256(entropy),
    info="contract_internal_secret", 
    length=32
  )

  // save to storage
  saveInternalSecretToStorage(internalSecret);

  // ...
}
\end{pseudocode}

\subsection{Notification seed algorithm}

Contracts derive per-recipient seeds using either custom shared secrets or deterministic generation from the internal secret.

\begin{pseudocode}
fun getSeedFor(recipientAddr) {
  // recipient has a shared secret with contract
  let seed := sharedSecretsTable[recipientAddr]

  // no explicit shared secret; derive seed using contract's secret
  if NOT exists(seed):
    seed := hkdf_sha256(
      ikm=contractInternalSecret,
      info=canonical(recipientAddr)
    )

  return seed
}
\end{pseudocode}

\subsection{Notification ID generation}

\begin{pseudocode}
fun notificationIdFor(contractOrRecipientAddr, channelId, env) {
  let salt := nil

  // depending on which mode the channel operates in
  if inCounterMode(channelId):
    // counter reflects the nth notification for the given 
    // contract/recipient in the given channel
    let counter := getCounterFor(contractOrRecipientAddr, channelId)
    salt := uintToDecimalString(counter)

  // otherwise, channel is in TxHash Mode or Bloom Mode
  else:
    salt := env.txHash

  // compute notification ID for this event
  let seed := getSeedFor(contractOrRecipientAddr)
  let material := concatStrings(channelId, ":", salt)
  let notificationId := hmac_sha256(
    key=seed,
    message=utf8ToBytes(material)
  )

  return notificationId
}
\end{pseudocode}

\subsection{Data Encryption and Decryption} Pseudocode for encrypting data into single-recipient notifications (contract) and decrypting data from single-recipient notifications (client).

\begin{pseudocode}
fun encryptNotificationData(
  recipientAddr,
  channelId,
  plaintext,
  env
) {
  // ChaCha20 expects a 96-bit (12 bytes) nonce, so 
  // combine two 12 byte buffers to create nonce
  let saltBytes := nil

  // depending on which mode the channel operates in
  if inCounterMode(channelId):
    // counter reflects the nth notification for the given recipient
    // in the given channel
    let counter := getCounterFor(recipientAddr, channelId)

    // encode uint64 counter in BE and left-pad with 4 bytes of 0x00
    // to make 12 bytes
    saltBytes := concat(zeros(4), uint64BigEndian(counter))

  // otherwise, channel is in TxHash Mode
  else:
    // take first 12 bytes of tx hash (make sure to decode the hex string)
    saltBytes := slice(hexToBytes(env.txHash), 0, 12)

  // take the first 12 bytes of the channel id's sha256 hash
  let channelIdBytes := slice(sha256(utf8ToBytes(channelId)), 0, 12)

  // produce the nonce by XOR'ing the two previous 12-byte results
  let nonce := xorBytes(channelIdBytes, saltBytes)

  // right-pad the plaintext with 0x00 bytes until it is of the desired
  // length (keep in mind, payload adds 16 bytes for tag)
  let message := concat(plaintext, zeros(DATA_LEN - len(plaintext)))

  // construct the additional authenticated data
  let aad := concatStrings(env.blockHeight, ":", env.txHash)

  // encrypt notification data for this event
  let seed := getSeedFor(recipientAddr)
  let [ciphertext, tag] := chacha20poly1305_encrypt(
    key=seed,
    nonce=nonce,
    message=message,
    aad=aad
  )

  // concatenate ciphertext and 16 bytes of tag
  // (note: crypto libs typically default to doing it this way in `seal`)
  let payload := concat(ciphertext, tag)

  return payload
}
\end{pseudocode}

\begin{pseudocode}    
fun decryptNotificationData(contractAddr, channelId, payload, env) {
  // depending on which mode the channel operates in
  if inCounterMode(channelId):
    // counter reflects the nth notification for the given recipient 
    // in the given channel
    let counter := getCounterFor(recipientAddr, channelId)

    // encode uint64 counter in BE and left-pad with 4 bytes of 0x00 
    // to make 12 bytes
    saltBytes := concat(zeros(4), uint64BigEndian(counter))

  // otherwise, channel is in TxHash Mode
  else:
    // take first 12 bytes of tx hash (make sure to decode the hex string)
    saltBytes := slice(hexToBytes(env.txHash), 0, 12)

  // ChaCha20 expects a 96-bit (12 bytes) nonce
  // take the first 12 bytes of the channel id's sha256 hash
  let channelIdBytes := slice(sha256(utf8ToBytes(channelId)), 0, 12)

  // produce the nonce by XOR'ing the two previous 12-byte results
  let nonce := xorBytes(channelIdBytes, counterBytes)

  // construct the additional authenticated data
  let aad := concatStrings(env.blockHeight, ":", env.txHash)

  // split payload
  let ciphertext := slice(payload, 0, len(payload) - 16)
  let tag := slice(payload, len(ciphertext))

  // decrypt notification data
  let seed := getSeedFor(contractAddr)
  let message := chacha20poly1305_decrypt(
    key=seed,
    nonce=nonce,
    message=ciphertext,
    tag=tag, 
    aad=aad
  )

  // do not trim trailing zeros because there is no END marker in CBOR. 
  // just decode plaintext as-is
  let plaintext := message

  return plaintext
}
\end{pseudocode}

\end{document}